# Design Considerations for 2D Dirac-Source FETs: Device Parameters, Non-Idealities and Benchmarking

Peng Wu, *Member, IEEE*, and Joerg Appenzeller, *Fellow, IEEE*

***Abstract*—Dirac-source field-effect transistors (DS-FETs) have been proposed as promising candidates for low-power switching devices by leveraging the Dirac cone of graphene as a low-pass energy filter. In particular, using two-dimensional (2D) materials as the channel in a DS-FET is of interest for ultimate scaling purposes. In this paper, we investigate the design considerations for 2D DS-FETs using ballistic simulations based on Landauer formalism. We study the impact of several key device parameters on the device performance, such as graphene doping, Schottky barrier heights, and effective mass of the 2D channel. In addition, we study the impact of non-idealities on the performance of DS-FETs, such as graphene disorder and rethermalization, as well as ways to mitigate them. Finally, we benchmark the performance of DS-FETs for different channel materials, providing a guide for the proper choice of material for 2D DS-FETs.**

## I. Introduction

SCALING down the supply voltage ($V_{DD}$) is critical for reducing the power consumption of transistors and integrated circuits (ICs). However, the subthreshold swing (SS) of conventional MOSFETs is fundamentally limited to ~60 mV/dec at room temperature, which hinders the reduction of $V_{DD}$, while maintaining a low off-state current ($I_{OFF}$). To overcome the SS limit, several novel device concepts have been proposed, such as tunnel field-effect transistors (TFETs) [1]–[5], negative-capacitance FETs (NC-FETs) [6], [7], nanoelectromechanical (NEM) relays [8] and impact-ionization MOS (IMOS) [9], [10]. In particular, Dirac-source field-effect transistors (DS-FETs) have been recently proposed and demonstrated as a promising candidate [11], [12]. DS-FETs leverage the Dirac cone of graphene as a low-pass energy filter to enable a "cold electron source", thus achieving a steep slope at room-temperature. Unlike TFETs, in which the transmission of band-to-band tunneling (BTBT) limits the on-state current $I_{ON}$, a DS-FET does not suffer from this limitation and has the potential of achieving a large $I_{ON}$ alongside with a steep slope.

The first demonstration of a DS-FET had been based on a carbon nanotube (CNT) as channel material [11]. However, there has also been interest in building DS-FETs using two-dimensional (2D) materials, such as molybdenum disulfide ($MoS_2$) [12]–[14], since the atomically thin body thicknesses of 2D materials allow for ultimate scaling of device dimensions. Moreover, integration becomes easier for 2D channels than individual 1D tubes. In this paper, we study the design considerations for 2D DS-FETs.

The paper is organized as follows. In Section II, we study the basic operation principles of 2D DS-FETs. In Section III, we discuss the impact of device parameters on the performance of 2D DS-FETs as well as the impact of non-idealities, based on ballistic simulations using Landauer formalism. Finally, in Section IV, we benchmark different channel materials for 2D DS-FETs, before concluding in Section V.

## II. Basic Operation Principles

In this section, we discuss the basic operation principles of 2D DS-FETs. Without loss of generality, we discuss n-type DS-FETs in this paper and the discussions apply to p-type DS-FETs by reversing the doping type of the graphene Dirac source (from p-type to n-type) and the polarities of gate and drain biases (from positive to negative), as well as by replacing the transport parameters (effective mass and valley degeneracy) of electrons with those of holes.

Fig. 1(a) and Fig. 1(b) show the schematic and band diagram of a 2D DS-FET. The device is composed of a graphene source, a 2D semiconductor channel (2DSC), source (S) and drain (D) contacts, a control gate (CG) that controls the doping in the graphene Dirac source, and a gate (G) that controls the channel as well as a segment of the graphene that makes contact to the 2DSC. For the n-type DS-FET under study, the graphene Dirac source is p-doped, while the graphene segment in contact with the 2DSC channel is n-doped by the gate G in the on-state to lower the Schottky barrier height to the 2DSC [15]–[18]. Note that the band movement of the graphene segment under control of G has not been depicted in Fig. 1(b) for simplicity, which we will reevaluate later in the manuscript. The basic operation of a 2D DS-FET can be understood from the Landauer formula [19]–[21]:

Manuscript received XXX, 2022; This work was supported by Intel Corporation.

Peng Wu was with the School of Electrical and Computer Engineering, Purdue University, West Lafayette, IN 47907 USA, and also with the Birck Nanotechnology Center, Purdue University, West Lafayette, IN 47907 USA. He is now with the Research Laboratory of Electronics, Massachusetts Institute of Technology, Cambridge, MA 02139 USA (e-mail: wu936@alumni.purdue.edu).

Joerg Appenzeller is with the School of Electrical and Computer Engineering, Purdue University, West Lafayette, IN 47907 USA, and also with the Birck Nanotechnology Center, Purdue University, West Lafayette, IN 47907 USA.

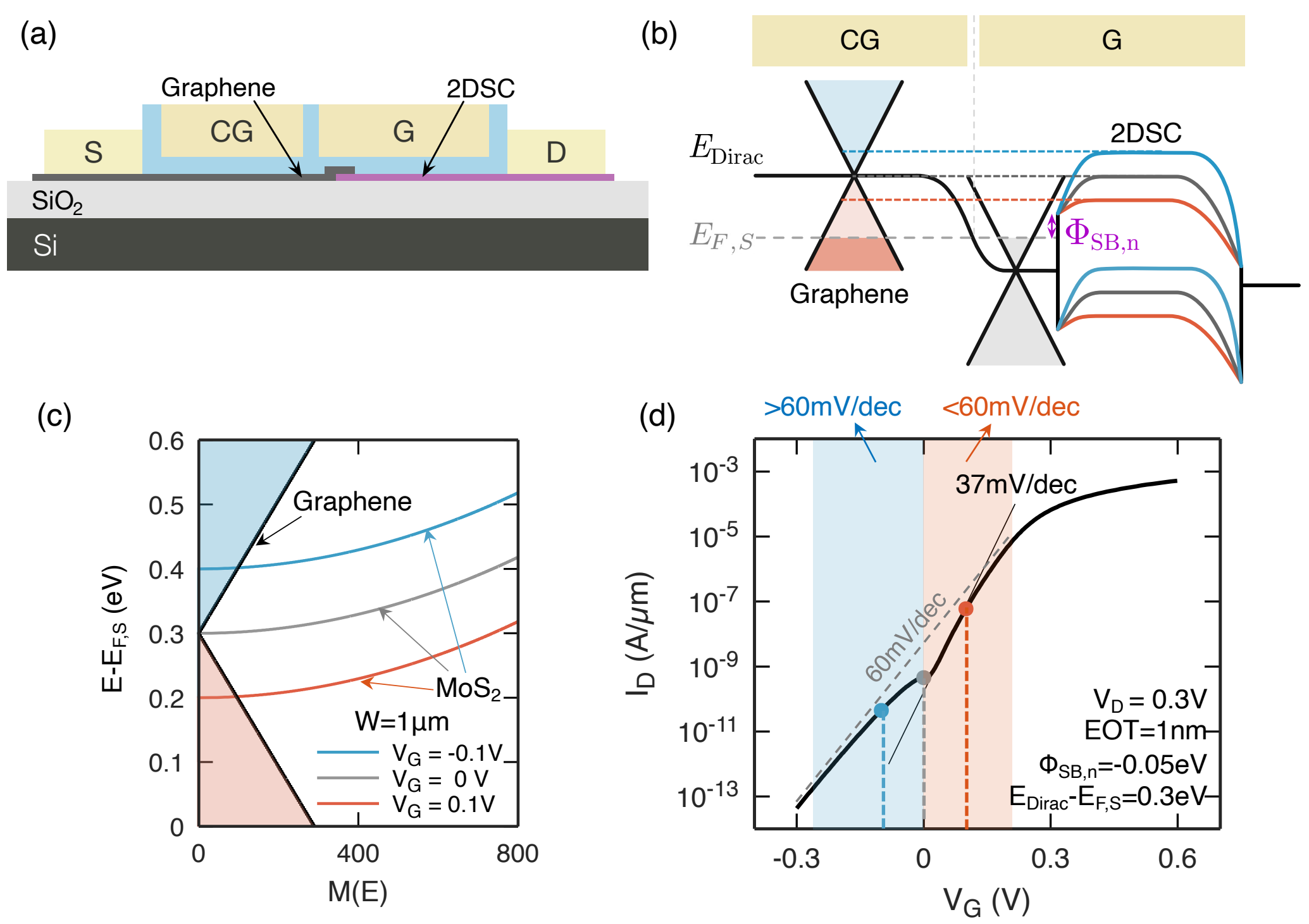

Fig. 1. Basic operation principle of a 2D DS-FET. (a) Schematic of a 2D DS-FET. (b) Band diagram of a 2D DS-FET at different gate voltages applied to G. (c) Number of modes of the graphene source and the $MoS_2$ channel at different gate voltages. (d) Simulated transfer characteristics of an $MoS_2$ DS-FET.

$$I=\frac{2q}{h}\int M(E)T(E)[f(E-E_{F,S})-f(E-E_{F,D})]\mathrm{d}E \quad (1)$$

where $T(E)$ is the transmission probability, $f$ is the Fermi-Dirac distribution function, and $M(E)$ is number of modes, which is determined by the smaller one of i) $M(E)$ in the graphene source and ii) $M(E)$ in the 2DSC channel, given by:

$$M_{\mathrm{Gr}}(E)=W\frac{2|E-E_{\mathrm{Dirac}}|}{\pi\hbar v_F} \quad (2)$$

$$M_{\mathrm{2DSC}}(E)=Wg_C\frac{\sqrt{2m_e^*(E-E_C)}}{\pi\hbar} \quad (3)$$

where $W = 1$ μm is the device width, $v_F = 10^6$ m/s is the Fermi velocity of graphene, $E_{\mathrm{Dirac}}$ is the energy at the Dirac point in the graphene source, $g_C$ is the conduction band valley degeneracy of the 2DSC, $m_e^*$ is the electron effective mass of the 2DSC, and $E_C$ is the conduction band minimum of the 2DSC channel. As an example, Fig. 1(c) shows the number of modes of the graphene source and the 2DSC channel, in this case, an $MoS_2$ channel ($m_e^*$ = 0.51 $m_0$, $g_C$ = 2 [22]). In general, the graphene source has a smaller $M(E)$ than the $MoS_2$ channel for energy levels above $E_C$ of $MoS_2$ (except very close to $E_C$). Therefore, $M(E)$ in the Landauer formula (1) is determined by $M_{\mathrm{Gr}}(E)$, while the $MoS_2$ channel behaves as a gate-tunable energy barrier that blocks the current below $E_C$ (here we ignore any ambipolar hole injection from the drain side due to a large bandgap of 1.8 eV of monolayer $MoS_2$ [23]). If we also assume that $T(E) = 1$ for $E > E_C$ (i.e. we ignore scattering in the $MoS_2$ channel and assume perfect transmissions for the Klein tunneling in the graphene p-n junction [24], [25] and at the graphene-$MoS_2$ contact) and that the drain voltage is large enough so that $f(E-E_{F,D}) \approx 0$ within the energy range of interest ($E > E_C$), we can rewrite (1) as:

$$I=\frac{2q}{h}\int_{E_C}^{\infty}M_{\mathrm{Gr}}(E)f(E-E_{F,S})\mathrm{d}E \quad (4)$$

Note that $M_{\mathrm{Gr}}(E)$ in the Dirac source is independent of gate voltage. Therefore, we can treat $M_{\mathrm{Gr}}(E) \times f(E-E_{F,S})$ as an effective distribution function, and thus $M_{\mathrm{Gr}}(E)$ filters the Fermi-Dirac tail and modulates the SS of the DS-FET. Since $f\approx\exp[-(E-E_{F,s})/k_BT]$ (for $E$-$E_{F,S} \gg k_BT$, where $k_B$ is Boltzmann constant, $T$ = 300 K is temperature) is a fast-decaying exponential function, a small energy range of ~$k_BT$ above $E_C$ contributes to most of the current $I$ in eq. (4) in the off-state of the DS-FET ($E_C$-$E_{F,S} \gg k_BT$). Therefore, $M_{\mathrm{Gr}}(E)$ near $E_C$ of the channel determines the SS of the device. As indicated by the orange lines in Fig. 1(b) and Fig. 1(c), if $E_C$ of $MoS_2$ aligns with the lower cone of the graphene source, $M_{\mathrm{Gr}}(E)$ decreases with increasing energy and the effective distribution function $M_{\mathrm{Gr}}(E) \times f(E-E_{F,S})$ is a super-exponential decaying function with energy $E$, thus the graphene source acts as a low-pass energy filter and enables a sub-60 mV/dec switching. However, if $E_C$ of $MoS_2$ crosses the Dirac point and reaches the upper cone of the graphene source, indicated by the blue lines in Fig. 1(b) and Fig. 1(c), $M_{\mathrm{Gr}}(E)$ increases with increasing energy, and thus the graphene source acts as a high-pass filter and the SS becomes larger than 60 mV/dec. The high-pass filtering above the Dirac point has a strong impact on the device characteristics and should not be ignored [13], [14] when analyzing DS-FET operation.

Next, we simulate the transfer characteristics of an $MoS_2$ DS-FET based on Landauer formula (1). In the simulation, we have considered the quantum capacitance $C_q$ [26] of $MoS_2$ to calculate the band movement of $E_C$ with gate voltage $V_G$:

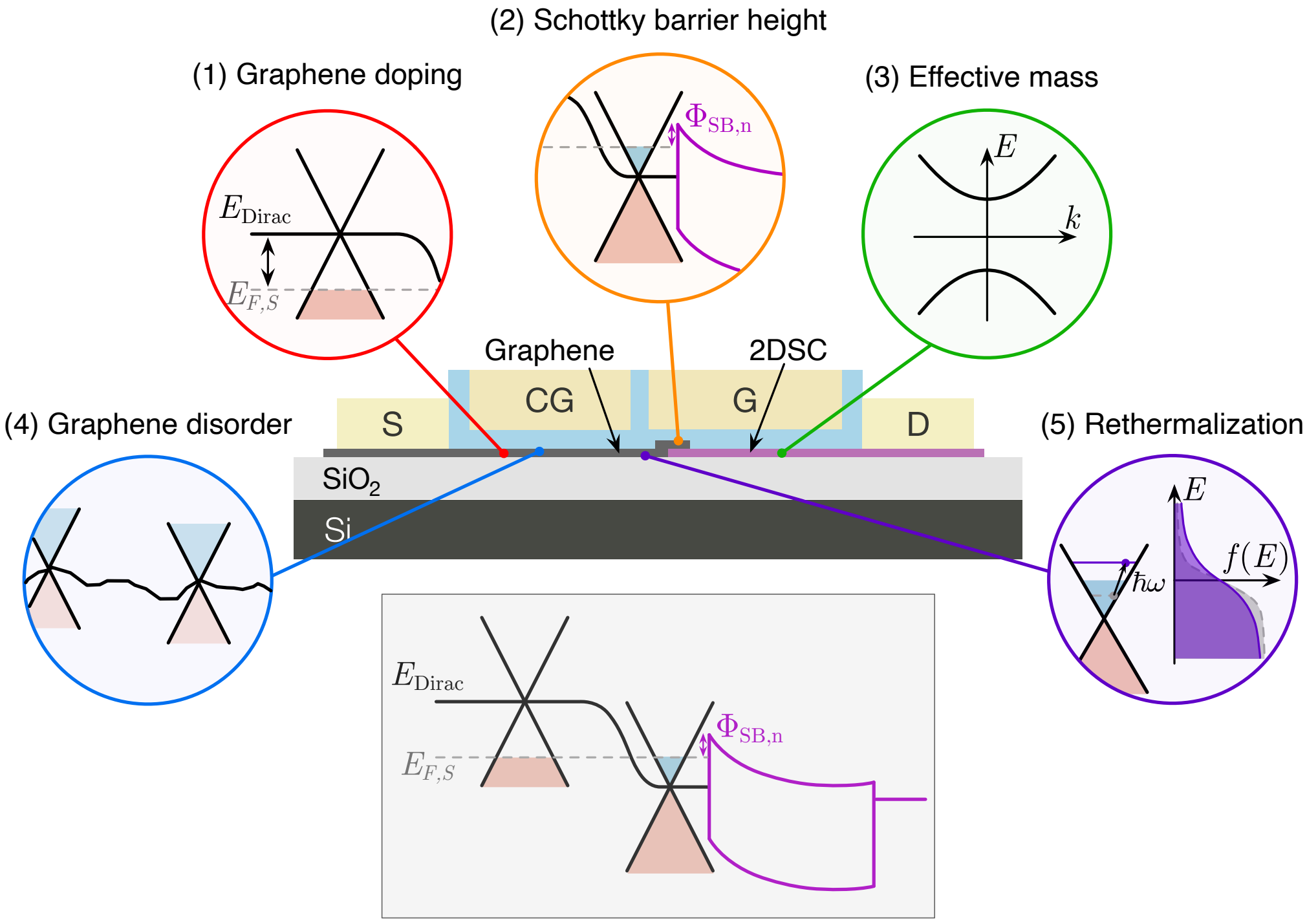

Fig. 2. Design considerations for 2D DS-FETs, including device parameters: (1) graphene doping; (2) Schottky barrier height; (3) effective mass; and non-ideal effects: (4) graphene disorder; (5) rethermalization.

$$\frac{d\psi_S}{dV_G} = \frac{d(-E_C/q)}{dV_G} = \frac{C_{\mathrm{ox}}}{C_{\mathrm{ox}}+C_q} \quad (5)$$

where $\psi_S$ is channel potential, $C_{ox}$ is gate oxide capacitance (we assume an EOT of 1 nm for the gate oxide). The quantum capacitance $C_q$ is given by:

$$C_q = q^2\frac{1}{2}D_{2\mathrm{D}}f(E_C-E_{F,S}) = q^2\frac{1}{2}\frac{g_C m_e^*}{\pi\hbar^2}\frac{1}{1+\exp\left(\frac{E_C-E_{F,S}}{k_BT}\right)} \quad (6)$$

where $D_{2D} = g_C m_e^*/\pi\hbar^2$ is the 2D density of states (DOS) of the $MoS_2$ channel, which is a constant value irrespective of energy in a 2D system, and the prefactor 1/2 accounts for the fact that only the positive velocity states are filled at large drain biases in a ballistic transistor [27]. We also set $E_C$-$E_{F,S}$ = 0.3 V at $V_G$ = 0 V as the origin for the band movement. Combined with eqs. (5) and (6), we can determine a one-to-one mapping between $V_G$ and $E_C$. Fig. 1(d) shows the simulated $I_D$ – $V_G$ transfer characteristics of an $MoS_2$ DS-FET. One can see that the DS-FET reaches an SS of 37 mV/dec, thanks to the aforementioned low-pass filtering effect. The SS value is in agreement with a previous simulation study on 2D $MoS_2$ DS-FETs [12]. In addition, as indicated by the grey dot in Fig. 1(d), a kink can be observed in the $I_D - V_G$ curve at $V_G = 0$ V, which separates the region with SS below 60 mV/dec [orange shaded region in Fig. 1(d)] and the region with SS above 60 mV/dec [blue shaded region in Fig. 1(d)]. The kink corresponds to where the $E_C$ of $MoS_2$ crosses the Dirac point of the graphene source, indicated by the dashed grey line in Fig. 1(b), which is in agreement with our previous discussion on the change from low-pass filtering below the Dirac point to high-pass filtering above the Dirac point. Such kink is a signature of DS-FET, and its existence can be used to validate the DS-FET operation. Also note that the steepest slope is located right above the kink in the $I_D - V_G$ curve, corresponding to when the $E_C$ of $MoS_2$ channel is just below the Dirac point in the graphene source. This is because $M_{\mathrm{Gr}}(E)$ is approaching zero close to the Dirac point and thus gives rise to the strongest low-pass filtering effect.

## III. Device Parameters and Non-Idealities

In this section, we discuss the impact of device parameters and non-idealities on the performance of 2D DS-FET. We discuss the following devices parameters: (1) graphene doping; (2) Schottky barrier height; (3) effective mass; and the following non-ideal effects: (4) graphene disorder; (5) rethermalization, as shown in Fig. 2.

### A. Graphene Doping

In the previous section, we have shown that the steepest slope occurs just above the kink in the $I_D$ – $V_G$ curve, which corresponds to the situation when the $E_C$ of $MoS_2$ is just below the Dirac point energy $E_{\mathrm{Dirac}}$ of the graphene source, as shown in Figs. 1(b-d). By tuning the doping levels in the graphene Dirac source using the control gate (CG), one can tune the energetic difference between $E_{\mathrm{Dirac}}$ and the source Fermi level $E_{F,S}$ and change the band position of $E_C$ (and thus the $V_G$ and $I_D$) at which the steepest slope is achieved.

Fig. 3 shows the simulated transfer characteristics of an $MoS_2$ DS-FET with different doping levels in the graphene Dirac source, expressed as $E_{\mathrm{Dirac}}$-$E_{F,S}$. One can see that with a higher doping level, the position of the kink in the $I_D$ – $V_G$ characteristics shifts to a more negative $V_G$ and a smaller $I_D$. Since the SS is not constant and has an abrupt change at the kink, rather than comparing the steepest SS at one point, a more appropriate way to evaluate the performance is to compare the $\Delta V_G$ (i.e. the $V_G$ sweep range) needed for the same $I_{OFF}$ and $I_{ON}$ (e.g., 1 nA/μm and 10 μA/μm, respectively). For a too high doping level ($E_{\mathrm{Dirac}}$-$E_{F,S}$ = 0.4 eV, blue line in Fig. 3), the steepest slope is achieved at a low current level below $I_{OFF}$ and the SS above $I_{OFF}$ is not as steep, which leads to a larger $\Delta V_G$. On the other hand, for a too low doping level ($E_{\text{Dirac}}$-$E_{F,S}$ = 0.2 eV, red line in Fig. 3), the kink is above $I_{OFF}$ and the SS between the kink and $I_{OFF}$ is above 60 mV/dec due to high-pass energy filtering above the Dirac point of graphene source, thus leading to a larger $\Delta V_G$. Moreover, in the on-state, $I_D$ at the same $V_G$ for $E_{\text{Dirac}}$-$E_{F,S}$ = 0.2 eV is decreased substantially compared with higher doping levels, since the $E_{\text{Dirac}}$ is closer to $E_{F,S}$ and there are fewer modes in the graphene that contributes to the current conduction. Thus a large $V_G$ is required to achieve the same $I_{ON}$, which leads to a further increase in $\Delta V_G$. Only for an optimal doping level ($E_{\text{Dirac}}$-$E_{F,S}$ = 0.3 eV, green line in Fig. 3), where the steepest SS is aligned with the target $I_{OFF}$, a lowest $\Delta V_G$ is achieved.

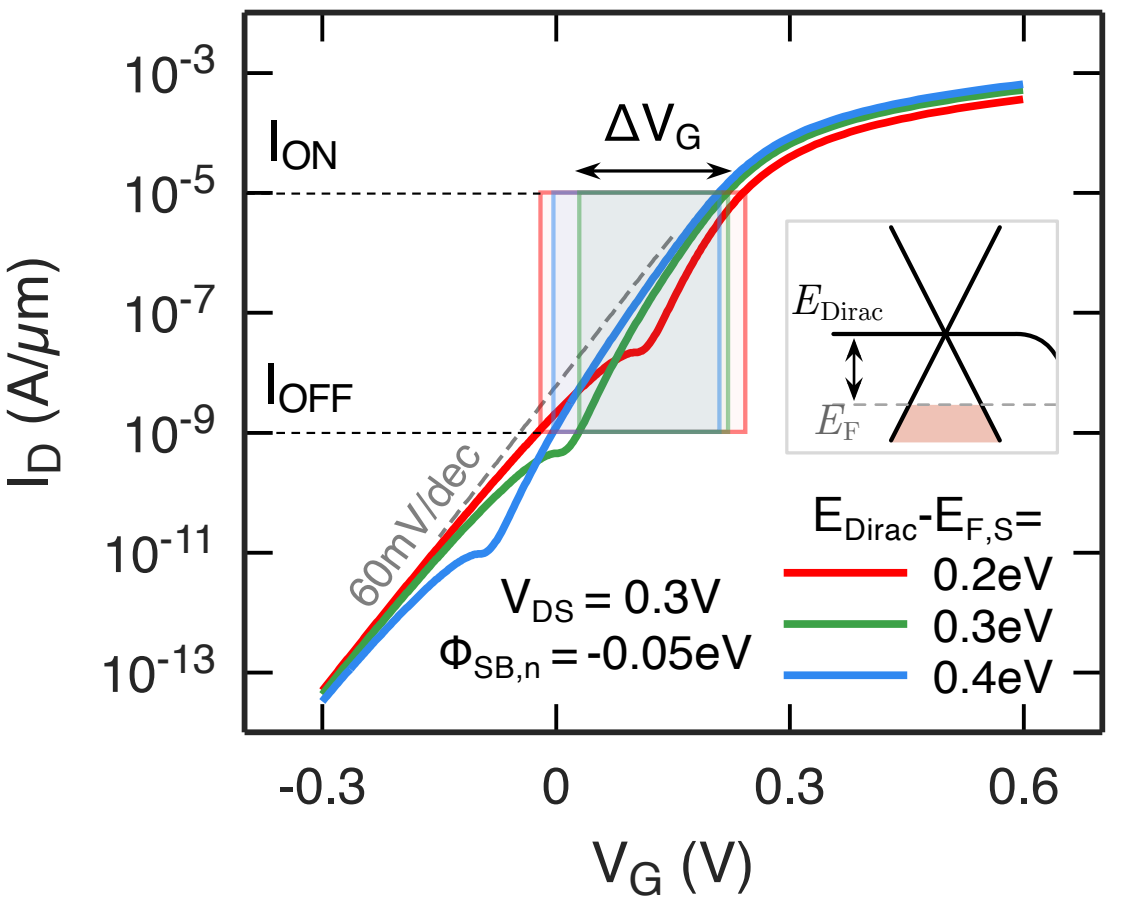

Fig. 3. Simulated transfer characteristics of an $MoS_2$ DS-FET with different doping levels in the graphene Dirac source.

In summary, the doping level of graphene affects the position of the kink in the $I_D - V_G$ characteristics, which is also the position of the steepest slope. Optimization of doping is essential to minimize $\Delta V_G$ of a DS-FET, which can be achieved by aligning the $I_D$ of the kink with the target off-current $I_{OFF}$. We later adopt this technique in Section IV for benchmark of DS-FETs. Moreover, the correlation between the doping level of graphene and the current level of the steep SS is another signature of DS-FET operation, which should be examined in previous experimental demonstrations of 2D DS-FETs [13], [14] to validate the DS-FET operation.

## *B. Schottky Barrier Height*

Next, we discuss the impact of the Schottky barrier height at the graphene-2DSC contact on the device performance of a 2D DS-FET. We show that a large Schottky barrier height not only leads to a lower on-state current, but may also degrades the SS.

In the previous simulations, we have assumed a negative Schottky barrier height [28] and a perfect transmission $T = 1$. However, for a positive Schottky barrier height, we need to consider the transmission of tunneling through a Schottky barrier, which can be calculated using Wentzel-Kramer-Brillouin (WKB) approximation [29], [30]:

$$T_{\text{WKB}}(E) = \exp\left(-2\int_{x_s}^{x_e} \kappa(E)\mathrm{d}x\right) \tag{7}$$

$$\kappa(E) = \frac{1}{\hbar}\sqrt{2m_e^*[E_C(x) - E]} \tag{8}$$

where $x_s$ and $x_e$ are the coordinates of the starting and ending points of the tunneling path, respectively, as shown in Fig. 4(a). With a larger Schottky barrier height, a lower transmission $T$ leads to a lower on-current. Meanwhile, from eqs. (7) and (8), we can see that electrons with a higher energy tunnel through a thinner barrier and thus have a higher transmission, i.e. $T_1 > T_2$ for $E_1 > E_2$, as illustrated in Fig. 4(a). This indicates that the Schottky barrier tunneling acts as a high-pass filter and degrades the SS [31], [32], which works against the low-pass filter in the graphene Dirac source. Therefore, a large Schottky barrier height may affect the steep slope switching of a DS-FET.

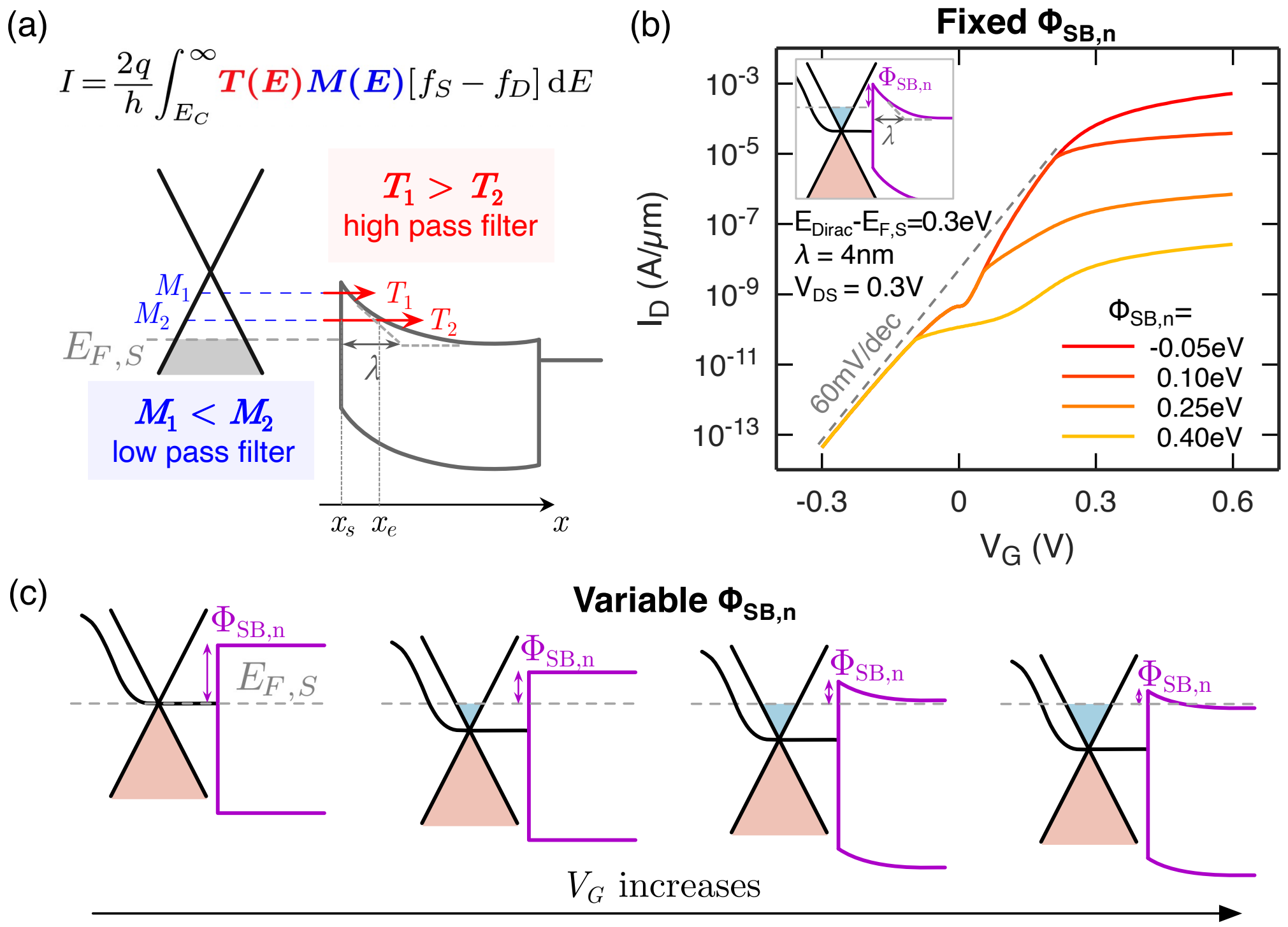

Fig. 4. (a) Illustration of Schottky barrier as a high pass filter. (b) Simulated transfer characteristics of an $MoS_2$ DS-FET with different Schottky barrier heights, assuming fixed Schottky barrier heights at graphene-$MoS_2$ contact. (c) Illustration of $V_G$-dependent Schottky barrier heights at graphene-$MoS_2$ contact.

Fig. 4(b) shows the simulated transfer characteristics of an $MoS_2$ DS-FET with different Schottky barrier heights for electrons, $\Phi_{SB,n}$. Note that in our device structure, since the gate G also covers a portion of the graphene that is in contact with the 2DSC channel [Fig. 1(a)], the Schottky barrier height is tuned by the gate voltage. Ignoring this effect for now, we have obtained the device characteristics as shown in Fig. 4(b). From Fig. 4(b), one can see that the on-current of the DS-FET decreases with a larger Schottky barrier height. Moreover, for a Schottky barrier height of 0.4 eV, the steep slope disappears because of the high-pass filtering effect of the large Schottky barrier.

According to Schottky-Mott rule [29], [33], the energy difference between the work function of undoped graphene (4.5 eV) and the electron affinity of $MoS_2$ (~4.1 eV) predicts a Schottky barrier height of ~0.4 eV. Thus, one may expect a degraded subthreshold slope for this device under any circumstances. However, as we mentioned before, the Schottky barrier height of the graphene-$MoS_2$ contact is tuned by the gate voltage. As illustrated in Fig. 4(c), at a small $V_G$, owing to the low DOS and thus the small $C_q$ of graphene near the Dirac point, the band movement of graphene is almost one-to-one with $V_G$ and is synchronized with that of $MoS_2$ [first and second panels in Fig. 4(c)]. Therefore, the Schottky barrier height decreases with increasing $V_G$ and no triangular shaped barrier is formed, thus avoiding the degradation of SS. When $V_G$ is further increased and $E_F$ is further spaced from $E_{\text{Dirac}}$, the DOS of graphene becomes larger and the band movement of graphene slows down, and only then a triangular barrier is formed [third panel in Fig. 4(c)]. Therefore, the Schottky barrier at this point, is expected to be substantially lower than 0.4 eV and a less severe degradation of SS than shown in Fig. 4(b) for constant $\Phi_{SB,n}$=0.4eV can be expected. For an even larger $V_G$, the Schottky barrier height continues to decrease when graphene becomes more n-doped [fourth panel in Fig. 4(c)], and an improved $I_{ON}$ is expected. In addition, previous experimental studies [16] and [17] have reported that even at zero gate voltage, the actual Schottky barrier height between graphene and $MoS_2$ is 100 to 150 meV, instead of 0.4 eV, due to charge transfer between graphene and $MoS_2$ upon contact, as indicated by DFT calculations [17].

In summary, for an $MoS_2$ DS-FET, an effective Schottky barrier height of ~100-150 meV is expected between the undoped graphene and $MoS_2$, which does not lead to a substantial degradation of the steep SS based on the simulation shown in Fig. 4(b); when the graphene is n-doped in the on-state ($V_G > 0$ V), the Schottky barrier height is further lowered by the n-type electrostatic doping of graphene, and an improved $I_{ON}$ can be expected. However, for other 2D materials, especially those with lower electron affinities than $MoS_2$ (~4.1 eV), such as $WS_2$ (3.96 eV) and $WSe_2$ (3.65 eV) [34], one needs to carefully evaluate the Schottky barrier height at the graphene-2DSC contact [35] to avoid degradation of SS and $I_{ON}$ of DS-FETs.

### C. Effective Mass of Channel Material

Next, we discuss the impact of the effective mass of the channel material on the performance of 2D DS-FETs. In particular, we show that a too small effective mass leads to an incomplete energy filtering and thus a degradation in SS of a DS-FET, while a too large effective mass leads to a limited band movement in the on-state and thus a degradation in the on-state current $I_{ON}$.

We study an InAs nanosheet ($m_e^* = 0.023\ m_0$, $g_C = 1$) and monolayer $MoS_2$ ($m_e^* = 0.51\ m_0$, $g_C = 2$) as the respective examples of channel materials with small and large effective masses. Note that although InAs is typically considered a 3D material, there have been various reports on using ultra-thin InAs nanosheets for transistor applications [36], [37], and thus we can use the 2D DOS of the lowest subband of the InAs nanosheet for the calculation and treat it as a 2D material. Also note that the small bandgap of InAs ($E_g = 0.354$ eV) may induce a pronounced ambipolar hole injection from the drain side, thus impacting the off-state current [30], [38]. Yet, to simplify the discussion, we ignore the hole injection from the drain and focus our discussion on the impact of the electron effective mass. Moreover, the quantization effect in the thickness direction leads to a larger bandgap of InAs nanosheets than its bulk value, thus alleviating the hole injection issue.

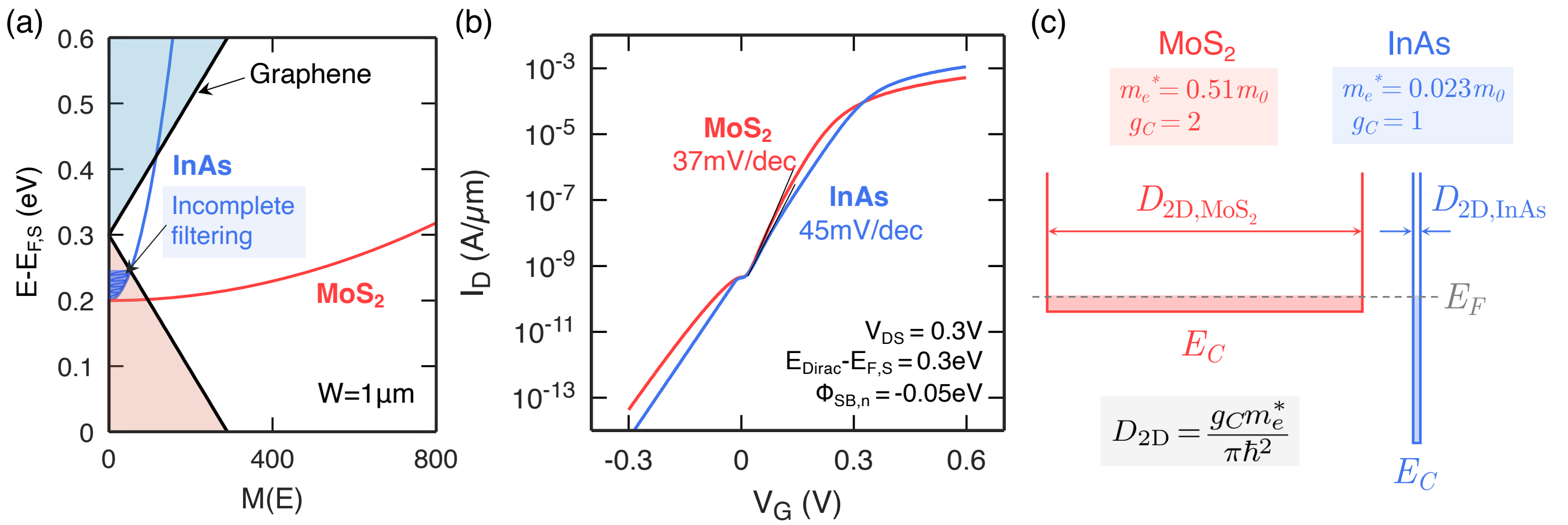

Fig. 5. (a) Number of modes in the graphene source, an InAs nanosheet and an $MoS_2$ channel, illustrating the incomplete filtering with an InAs channel. (b) Simulated transfer characteristics of $MoS_2$ and InAs DS-FETs. (c) Illustration of band movement in the ON-state in $MoS_2$ and InAs DS-FETs.

Fig. 5(a) shows the number of modes $M(E)$ of the graphene source (black line) and an InAs nanosheet channel (blue parabola), in comparison with $M(E)$ of an $MoS_2$ channel (red parabola). The InAs channel has a much lower $M(E)$ than the $MoS_2$ due to a much smaller effective mass ($m_e^* = 0.023\ m_0$ vs. $m_e^* = 0.51\ m_0$) and a smaller band degeneracy ($g_C = 1$ vs. $g_C = 2$). In an $MoS_2$ DS-FET, as we have discussed before, the number of modes is limited by the graphene Dirac source, and thus the low-pass energy filtering leads to a steep SS. However, in an InAs DS-FET, due to the smaller $M(E)$ of the InAs channel, the number of modes is limited by the $M(E)$ of InAs in the lower part of the energy window [the blue shaded region between the blue parabola and black line in Fig. 5(a)]. Therefore, since $M(E)$ does not decrease monotonically with higher energy $E$ in the energy window, the low-pass filtering is incomplete, and thus a degradation of SS is expected.

Fig. 5(b) shows the simulated transfer characteristics of an InAs DS-FET and an $MoS_2$ DS-FET. One can see that the SS of the InAs DS-FET deteriorates to 45 mV/dec due to the incomplete energy filtering, compared to 37 mV/dec in the $MoS_2$ DS-FET. Therefore, at low gate voltages, the drain current $I_D$ of the $MoS_2$ DS-FET is higher than that of the InAs DS-FET due to the steeper SS. However, at larger gate voltages, the drain current $I_D$ of the InAs DS-FET surpasses that of the $MoS_2$ DS-FET. The lower $I_{ON}$ of the $MoS_2$ DS-FET at larger gate voltages is the result of a limited band movement due to the larger effective mass of $MoS_2$. As illustrated in Fig. 5(c), the larger effective mass $m_e^*$ (and larger band degeneracy $g_C$) leads to a larger 2D DOS $D_{2D}$ (and thus a larger quantum capacitance $C_q$) for $MoS_2$. This in turn implies according to eqs. (5) and (6) that for the same gate voltage in the on-state, the conduction band $E_C$ moves much less below $E_F$ in case of $MoS_2$ than for InAs, as illustrated in Fig. 5(c). Therefore, the energy window for current conduction (roughly from $E_C$ to $E_F$ in the on state) is smaller in $MoS_2$, which leads to a degraded $I_{ON}$. Clearly, there is a $V_{DD}$ dependence on which material is more favorable for higher $I_{ON}$, which will be discussed further in the "Benchmark" Section IV.

Moreover, DS-FETs are even more susceptible to $I_{ON}$ degradation induced by large effective masses than MOSFETs. In conventional MOSFETs, the smaller energy window for a material with a large effective mass can be compensated by a larger number of modes $M(E)$, i.e., a larger integrand compensates for a smaller range of integration in the Landauer formula [eq. (1)]. However, in DS-FETs, since the graphene Dirac source is the bottleneck for $M(E)$, a material with a large effective mass does not benefit from a large $M(E)$, yet still suffers from a smaller energy window. Therefore, materials with large effective masses are at an even larger disadvantage in DS-FETs than in MOSFETs in terms of $I_{ON}$.

In summary, the effective mass of the channel material has a strong impact on the characteristics of a DS-FET. A too small effective mass leads to a degraded SS due to an incomplete low-pass filtering, while a too large effective mass leads to a degradation in $I_{ON}$ due to a limited band movement in the on-state, which is further worsened by the $M(E)$ bottleneck of the Dirac source. Therefore, there exists an optimal effective mass for DS-FETs, which we will further explore in the "Benchmark" Section IV.

## *D. Graphene Disorder*

In the previous discussion, the graphene Dirac source is assumed to be disorder-free. However, it has been reported [39] that graphene is subject to random potential fluctuations caused by an inhomogeneous distribution of charged impurities in the substrate, leading to fluctuations of the Dirac point energy, as illustrated in Fig. 6(a). This effect, also known as "electron-hole puddles", results in a minimum conductivity of $4e^2/h$ of graphene [39]–[41], instead of the theoretical value of $4e^2/\pi h$ predicted by evanescent transport [42], a phenomenon often referred to as "the mystery of the missing $\pi$". Next, we discuss the impact of such disorder on the performance of a 2D DS-FET.

Fig. 6(b) shows the number of modes in a disordered graphene source and an $MoS_2$ channel. We model the disorder as a fluctuation of the Dirac point energy that follows a Gaussian distribution with a variance $\sigma = 50$ meV [39] and assume that the number of modes is a superposition of the $M(E)$ across the distribution of Dirac point energies. Compared with the $M(E)$ in an ideal graphene in Fig. 1(c), $M(E)$ in a disordered graphene exhibits a non-zero value at the charge neutrality point, and thus the low-pass filtering effect is deteriorated. Fig. 6(c) shows the impact of the disorder on the transfer characteristics of an $MoS_2$ DS-FET. With a disordered graphene source, the SS of the DS-FET degrades to 47 mV/dec due to a deteriorated low-pass filtering effect, compared to 37 mV/dec with an ideal graphene source.

We would like to point out that the energy variance $\sigma = 50$ meV in ref. [39] was extracted when graphene is close to the charge neutrality point, where the DOS at the Fermi level is the lowest. However, in the p-doped graphene Dirac source of a n-type DS-FET, the DOS at $E_{F,S}$ (thus the quantum capacitance $C_q$) is larger than at the charge neutrality point. Since the origin of the disorder is the inhomogeneities of charged impurities on the substrate, the larger $C_q$ results in a smaller fluctuation in $E_{\mathrm{Dirac}}$, i.e., $\sigma$ is expected to be lower than 50 meV in the p-doped graphene Dirac source, and thus the deterioration of SS is expected to be less severe than the prediction in Fig. 6(c). Moreover, since the length scale of the electron-hole puddles is ~30 nm [39], the impact of graphene disorder is further diminishing for DS-FETs with scaled dimensions.

Having discussed the performance degradation of a DS-FET due to an increased $M(E)$ close to the Dirac point that is caused by disorder, now we consider the performance improvement from a reduced $M(E)$ close to the Dirac point by introducing a bandgap in the graphene Dirac source, as illustrated in Fig. 6(d). A bandgap opening up to 250 meV has been reported in bilayer graphene by applying a vertical electric field [43], [44]. For simplicity, we assume an $E - k$ dispersion relation of:

$$E = E_{\mathrm{Dirac}} \pm \hbar v_F \sqrt{k^2 + (E_{g,\mathrm{DS}}/2)^2} \tag{9}$$

for the gapped Dirac source, where $E_{g,\mathrm{DS}}$ is the bandgap. $M(E)$ of the gapped graphene is calculated as:

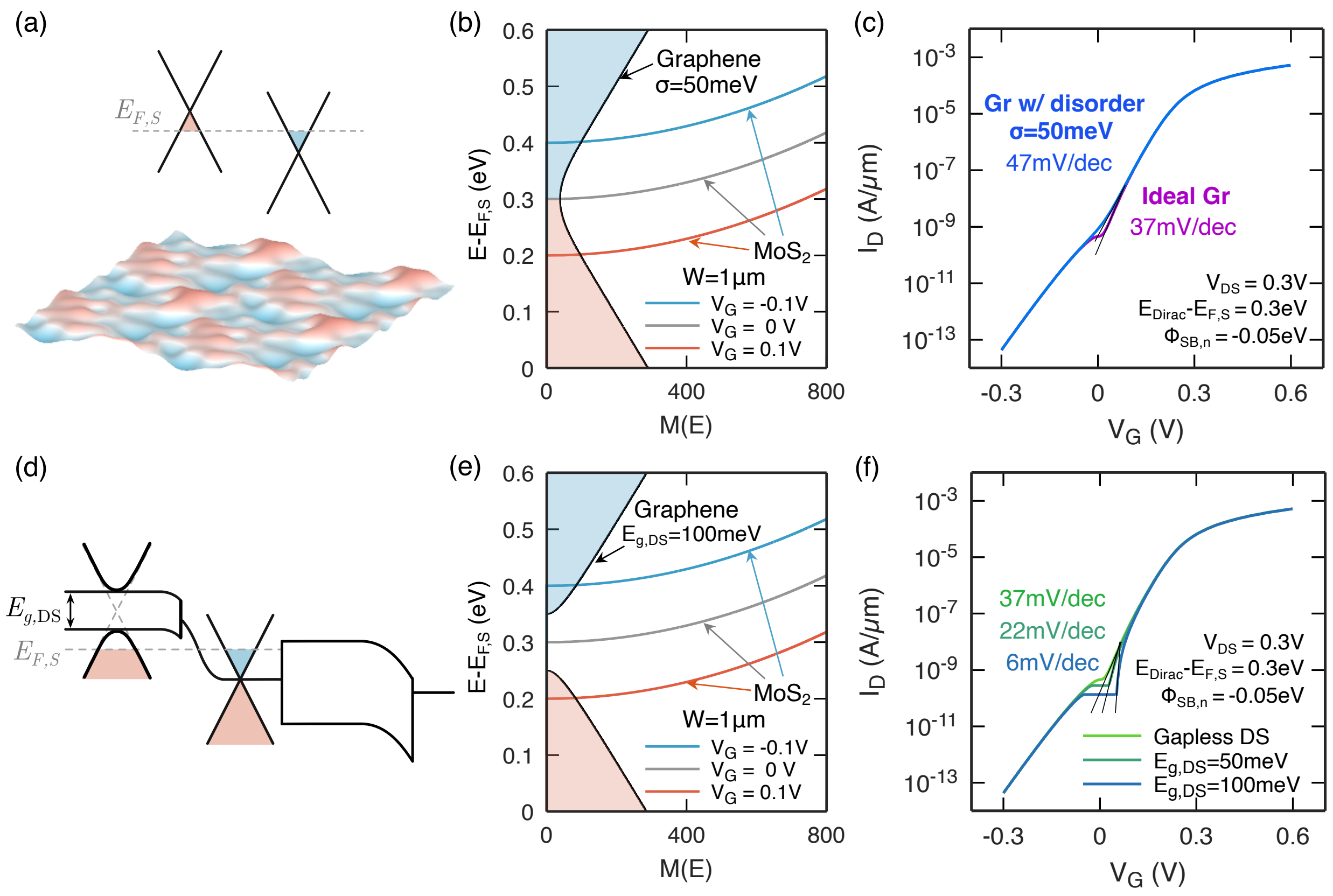

Fig. 6. (a) Illustration of electron-hole puddles in graphene. (b) Number of modes in a disordered graphene source and an $MoS_2$ channel. (c) Simulated transfer characteristics of $MoS_2$ DS-FETs with a disordered graphene source and an ideal graphene source. (d) Illustration of a DS-FET with a gapped Dirac source. (e) Number of modes in a gapped graphene Dirac source and an $MoS_2$ channel. (f) Simulated transfer characteristics of $MoS_2$ DS-FETs with different bandgaps in the Dirac sources.

$$M_{\mathrm{Gr,gapped}}(E)=\begin{cases} W\dfrac{2\sqrt{|E-E_{\mathrm{Dirac}}|^2-(E_{g,\mathrm{DS}}/2)^2}}{\pi\hbar v_F} \quad \text{for } |E-E_{\mathrm{Dirac}}|>E_{g,\mathrm{DS}}/2 \\ 0 \quad\quad \text{for } |E-E_{\mathrm{Dirac}}|\leqslant E_{g,\mathrm{DS}}/2 \end{cases} \tag{10}$$

which becomes eq. (2) for $E_{g,\mathrm{DS}}=0$. Fig. 6(e) shows the number of modes in a gapped graphene Dirac source with a bandgap of $E_{g,\mathrm{DS}}$ = 100 meV and an $MoS_2$ channel in a DS-FET. Due to a reduced $M(E)$ close to the Dirac point, a steeper SS is expected due to the stronger energy filtering effect. Fig. 6(f) shows the simulated transfer characteristics of DS-FETs with different bandgaps in the Dirac sources. The SS values are further improved to 22 mV/dec and 6 mV/dec for bandgap values of 50 meV and 100 meV, respectively, which are in reasonable agreement with a previous simulation study [12].

In summary, graphene disorder with a Dirac point energy variance of $\sigma$ = 50 meV results in a degradation of SS from 37 mV/dec to 47 mV/dec. However, since the disorder is smoothed out by the larger $C_q$ in the doped graphene source, the graphene disorder is not expected to become a major issue for a DS-FET. Moreover, we have evaluated the impact of a bandgap in the Dirac source, which resulted in an improved SS of 6 mV/dec for a bandgap of 100 meV.

## *E. Rethermalization*

In the previous discussion, we have ignored scattering and assumed ballistic transport conditions. Next, we show that scattering could lead to rethermalization of the cold carriers from the Dirac source and a degradation of SS back to 60 mV/dec.

Fig. 7 illustrates the rethermalization of the cold carriers from the Dirac source. As discussed in Section II, the origin of the steep slope switching in an n-type DS-FET is the low-pass filtering from the $M(E)$ in the p-type graphene Dirac source, which leads to a cold electron injection. In order for the cold electrons to contribute to a steep SS, they need to preserve their "coldness", i.e., a distribution function with an effective temperature colder than room temperature, before reaching the top-of-the-barrier (ToB) [27] in the channel of the 2D DS-FET. However, note that there is an n-doped segment of graphene in the DS-FET [Fig. 1(a)], which serves to lower the Schottky barrier height between graphene and 2DSC. Since the cold electrons injected from the p-type graphene source are not in thermal equilibrium with the n-type graphene, inelastic scattering in the n-doped graphene segment will destroy the $M(E)$ information carried over from the cold electron source and rethermalize the electrons [45], [46]. In the worst case scenario, the electron distribution is rethermalized to a room-temperature (RT) distribution, and the SS of the DS-FET goes back to 60 mV/dec (or even worse, considering the high-pass filtering effect of the n-doped graphene).

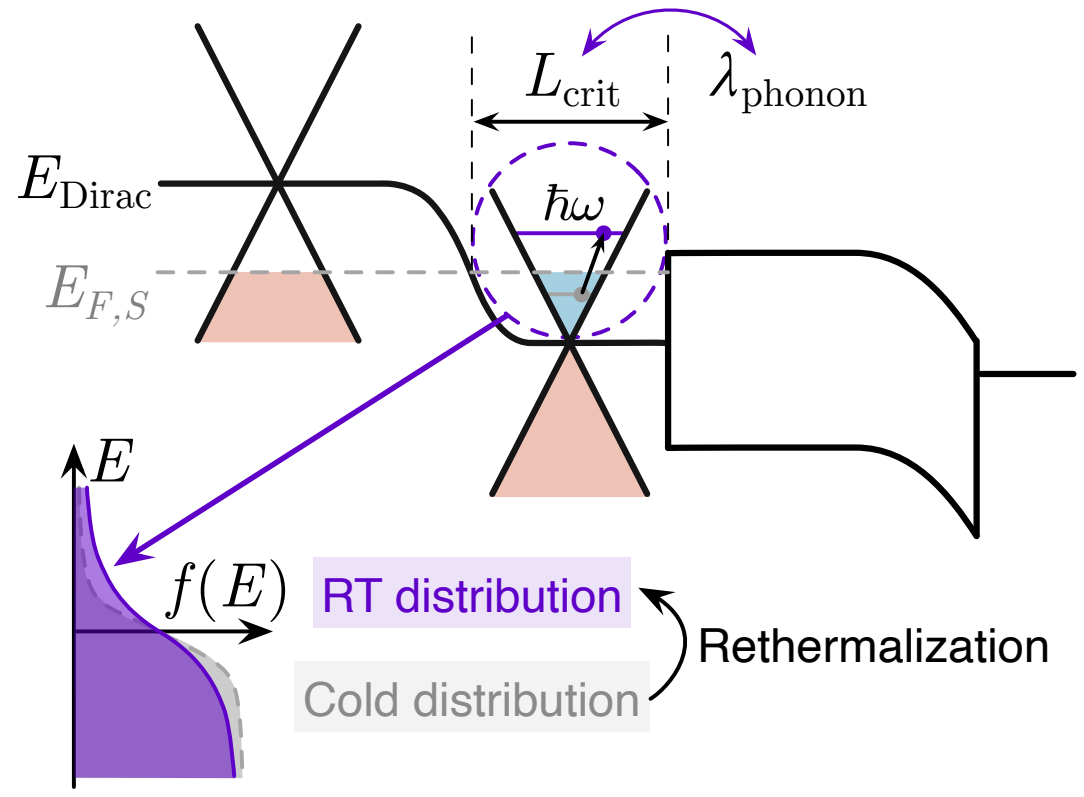

Fig. 7. Illustration of rethermalization of cold carriers due to phonon scattering.

To prevent the SS degradation induced by rethermalization, the length of the n-doped graphene segment should be short enough to minimize the scattering. Note that only inelastic scattering, such as phonon scattering, leads to rethermalization, while elastic scattering, such as charged impurity scattering, only changes the electron momentum but does not alter the energy spectrum, thus not contributing to rethermalization. Therefore, the length of the n-doped graphene segment, i.e., the critical length scale for rethermalization $L_{\text{crit}}$, should be much smaller than the mean free path associated with phonon scattering, $\lambda_{\text{phonon}}$, instead of the mean free path that accounts for all scattering events. A previous paper [47] has mapped out the electron mean free paths in graphene using scanning capacitance microscopy and analyzed the individual contributions to the mean free path from different scattering mechanisms, including resonant scatterers (RS), charged impurities (CI) and surface polar phonons (SPP). It was determined that the mean free path associated with SPP scattering is around 300 nm to 1 μm, depending on the substrate properties, such as permittivity [48]. Interestingly, previously demonstrated $MoS_2$ DS-FETs [13], [14] have rather long (5-10 μm) graphene-$MoS_2$ contact lengths. In light of the expected rethermalization described here, the occurrence of steep slopes in these devices is unexpected and should be carefully revisited to better understand their origin.

## IV. Benchmark

In this section, we discuss the benchmark of channel materials for 2D DS-FETs. We adopt a similar benchmarking methodology as in ref. [5], i.e., fixing an $I_{OFF}$ target and comparing the $I_{ON}$ for a given $V_{DD}$. The detailed procedures for the benchmarking effort, as illustrated in Fig. 8(a), is as follows:

i) Select the optimal doping level of the graphene Dirac source for a given $I_{OFF}$ target, i.e., align the kink in the $I_D - V_G$ characteristics with $I_{OFF}$;

ii) Shift the kink to $V_G = 0$ V (in real devices, this can be achieved with a proper $V_{th}$ tuning);

iii) Determine the $I_{ON}$, i.e., $I_D$ at $V_G = V_{DD}$.

For simplicity, we only consider the differences in electron effective mass $m_e^*$ and conduction band valley degeneracy $g_C$ for different channel materials in the benchmark, and assume other parameters to be the same, such as Schottky barrier height at the graphene-2DSC contact and EOT of the gate oxide, although these parameters can be material-dependent in real devices. Also, as in Section III-C, we ignore any potentially existing the ambipolar injection from the drain side.

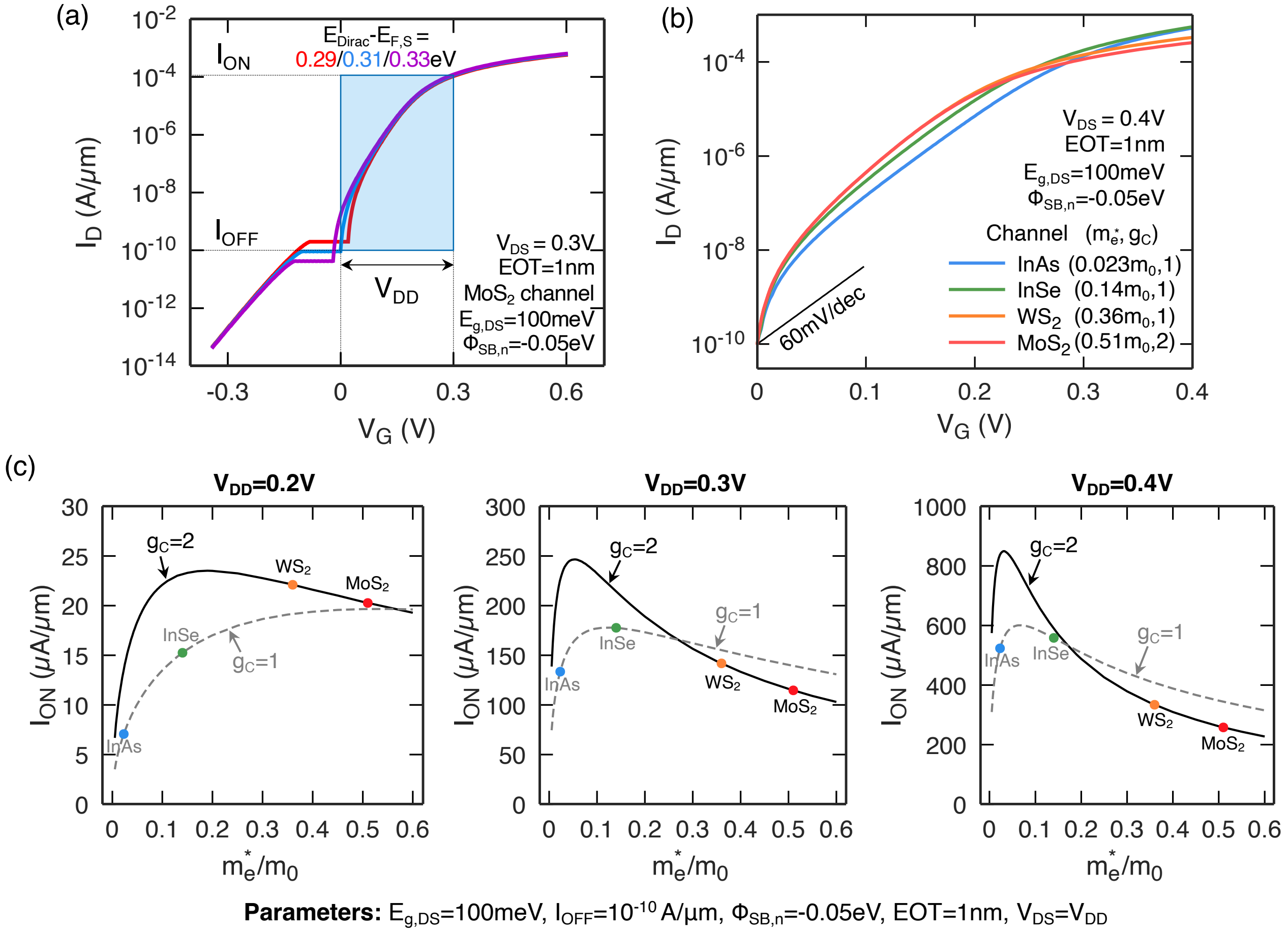

Fig. 8. (a) Illustration of the benchmarking methodology. (b) Simulated transfer characteristics of DS-FETs with different channel materials. (c) Benchmark of DS-FETs with different channel materials for $V_{DD}$ = 0.2 V, 0.3V and 0.4 V.

Fig. 8(b) shows the simulated transfer characteristics of DS-FETs for four different channel materials after doping level optimization and $V_{th}$ shifting: (a) InAs ($m_e^* = 0.023\ m_0$, $g_C = 1$); (b) InSe ($m_e^* = 0.14\ m_0$, $g_C = 1$) [49]; (c) $WS_2$ ($m_e^* = 0.36\ m_0$, $g_C = 2$) [50]; (d) $MoS_2$ ($m_e^* = 0.51\ m_0$, $g_C = 2$) [22]. An $I_{OFF}$ target of $10^{-10}$ A/μm is selected and a gapped graphene Dirac source with $E_{g,\mathrm{DS}}$ = 100 meV is used. As we have discussed in Section III-C, InAs and InSe DS-FETs exhibit degraded SS due to an incomplete low-pass filtering. Interestingly, even the on-currents are lower than for $WS_2$ and $MoS_2$ DS-FETs at $V_G = 0.2$ V because the devices are still deep in the off-state for this voltage. However, at higher $V_G$, because of the larger energy window for conduction with a smaller effective mass and band degeneracy, $I_{ON}$ of InAs and InSe DS-FETs surpass $I_{ON}$ of $WS_2$ and $MoS_2$ DS-FETs, with a crossover at around $V_G = 0.25$ V.

Fig. 8(c) shows the benchmark results, in terms of $I_{ON}$ as a function of effective mass of channel material, for DS-FETs at different supply voltages $V_{DD}$. For $V_{DD} = 0.2$ V, the peak $I_{ON}$ is achieved at $m_e^* \sim 0.18\ m_0$ and $g_C = 2$, and $WS_2$ has the best performance of the four materials under consideration, while InAs has the worst because of the SS degradation. For $V_{DD} = 0.3$ V, the peak $I_{ON}$ is achieved at $m_e^* \sim 0.05\ m_0$ and $g_C = 2$, and of the four materials under study, InSe has the highest $I_{ON}$, while $MoS_2$ has a lowest $I_{ON}$ due to a limited band movement. For $V_{DD} = 0.4$ V, the peak $I_{ON}$ is achieved at $m_e^* \sim 0.025\ m_0$ and $g_C = 2$, and of the four materials that are evaluated here, InSe has the best performance, and InAs comes as a close second, while $MoS_2$ again has the worst performance due to a limited band movement in the on-state. The general trend is that at a low $V_{DD}$, channel materials with a moderate $m_e^*$ that is not too small to exhibit SS degradation, nor too large to be limited in band movement, are more favorable for highest performance; while at a high $V_{DD}$, the optimal $m_e^*$ shifts to an even smaller value of $m_e^*$ for which SS degradation is already present, since the advantage of a larger band movement overcomes the disadvantage of a worse SS.

Finally, it is worth noticing that some factors have been ignored for convinience in our benchmarking, which may be important and should be taken into account for real applications. For example, while $WS_2$ outperforms $MoS_2$ at all $V_{DD}$ values in Fig. 8(c), the Schottky barrier height between graphene and $WS_2$ is expected to be larger than for $MoS_2$, which may degrade the performance. Another example is that the small bandgap of InAs ($E_g$ = 0.354 eV) may cause ambipolar injection from the drain and lead to a degradation of $I_{OFF}$, which we have ignored in the discussion.

## V. Conclusion

In this paper, we have discussed various design considerations of 2D DS-FETs. The basic operation principles of 2D DS-FETs are introduced, followed by a discussion on the impact of device parameters and non-idealities on the performance of 2D DS-FETs. Finally, we have benchmarked the performance of DS-FETs from different channel materials, such as InAs, InSe, $WS_2$ and $MoS_2$. Our study provides guidance for the design of 2D DS-FETs.